# Co/Ni multilayers for spintronics: high spin-polarization and tunable magnetic anisotropy

S. Andrieu[1*], T. Hauet[1], M. Gottwald[1,**], A. Rajanikanth[1,***], L. Calmels[2], A.M. Bataille[3], F. Montaigne[1], S. Mangin[1], E. Otero[4], P. Ohresser[4], P. Le Fevre[4], F. Bertran[4], A. Resta[4], A. Vlad[4], A. Coati[4], Y.Garreau[4,5]

[1] *Institut Jean Lamour, UMR CNRS 7198, Université de Lorraine, 54506 Vandoeuvre lès Nancy, France*
[2] *CEMES-CNRS, Université de Toulouse, BP 94347, F-31055 Toulouse Cedex 4, France*
[3] *Laboratoire Léon Brillouin, IRAMIS, CEA Saclay, 91191 Gif sur Yvette, France*
[4] *Synchrotron SOLEIL-CNRS, L'Orme des Merisiers, Saint-Aubin, BP48, 91192 Gif-sur-Yvette, France*
[5] *Laboratoire Matériaux et Phénomènes Quantiques, Université Paris Diderot, 75013 Paris, France*

In this paper we analyze in details the electronic properties of (Co/Ni) multilayers, a model system for spintronics devices. We use magneto-optical Kerr (MOKE), spin-polarized photoemission spectroscopy (SRPES), x-ray magnetic circular dichroism (XMCD) and anomalous surface diffraction experiments to investigate the electronic properties and perpendicular magnetic anisotropy (PMA) in [Co(x)/Ni(y)] single-crystalline stacks grown by molecular beam epitaxy. The magnetization transition from in-plane to out-of-plane was studied by MOKE varying the Ni coverage on Co in the sub-monolayer range, confirming definitely the interface origin of PMA in this system. Surprisingly the spin-polarization for Co terminated stacks is found to be much larger than bulk Co, reaching at least 90 % for 2 Co atomic planes. Angle-dependent XMCD using strong applied magnetic field allows us to show that the orbital moment anisotropy in Co is responsible for the PMA and that our results are consistent with Bruno's model. Finally, we checked by surface diffraction that the fcc stacking is preferred for 1ML Co–based superlattices, whereas the hcp stacking dominates for larger Co thicknesses. We finally checked the stacking influence on Co and Ni magnetic moments by using *ab initio* calculations.

\* corresponding author : stephane.andrieu@univ-lorraine.fr
\*\* new address : IBM - T. J. Watson Research Center, Yorktown Heights, New York 10598, USA
\*\*\*new address : University of Hyderabad, Hyderabad 500046. Telangana. INDIA.

Research on spintronic devices such as magnetic random access memories (MRAM) and magnetic sensors have generated the last decades a perpetual need for original magnetic materials. Although today most of spintronic devices are based on a similar spin-valve magnetic stack for sensing, writing and reading parts, each application requires specific optimized features for the magnetic films. For instance, in the view of Spin Transfer Torque (STT)-MRAM [SANK07, MEEN14] and STT-oscillators [SANK07, SILV10] implementation, magnetic tunnel junction (MTJ) with thin electrodes having low damping, high perpendicular anisotropy, high spin-polarization and moderate magnetization are investigated. Different ways are heavily pursued. One involves rare-earth/transition metal ferrimagnet alloys [DUME12]. Another one is [Fe$_{1-x}$Co$_x$/Pd] and [Fe$_{1-x}$Co$_x$/Pt] multilayers [JOHN96, ANDE06, OUAZ12]. Both show large perpendicular magnetocrystalline anisotropy (PMA) and allow easy tuning of magnetization. However these systems offer only a low spin-polarization and high damping parameter [RAJA10]. On the contrary Heusler alloys are heavily studied since they are theoretically 100% spin-polarized with extremely low damping (<0.001). High Tunnel MagnetoResistance (TMR) ratio have been indeed reported for magnetic tunnel junctions with Co-based full Heusler alloy electrodes [SAKU06, MARU06, TEZU06] and damping below 0.001 was reported [ANDR16, PRAD17] but large PMA is still challenging in this kind of materials.

Besides [Co/Ni] multilayers have gained large interest in view of spin-transfer applications since it may provide all the requested features cited above [YOU12, GOTT12]. The Co/Ni interface produces a magnetic anisotropy that is perpendicular to the interface [JOHN92, ZHAN93, GIRO09]. PMA up to 5 MJ/m$^3$ can be achieved for superlattices of 1 MonoLayer (ML) of Co and 3 ML of Ni grown along the (111) direction [JOHN96, GOTT12, SHAW13]. Changing the thickness of Co allows an easy tuning of PMA amplitude and calculations lead to similar conclusions [DAAL92, KYUN96, GIMB12]. Magnetization of Co/Ni multilayers is moderate (around 700 emu/cm$^3$ for Co 1ML / Ni 3ML). Gilbert damping has been found to be quite insensitive to the composition and can reach values around 0.02 [BEAU07, CHEN08, SEKI17]. The importance of such a set of characteristics to enhance spin-tranfer has been recently demonstrated. First reliable STT switching has been achieved in fully metallic Co/Ni-based GMR nanopillars for low critical current [MANG06, MANG09] and for sub-nanosecond time [BERN11]. Second recent reports on STT-induced domain wall motion have demonstrated low critical current [TANI09] and high domain wall speed in Co/Ni [UEDA12, LEGA15, LEGA17]. Finally, a high Spin Polarization (denoted as SP in the following) was predicted using *ab initio* calculation [GIMB11] and indirectly estimated experimentally [UEDA12]. The main problem to use Co/Ni in magnetic tunnel junction is to find a suitable insulating barrier for getting large tunnel magnetoresistance [YOU12]. Encouraging results were recently reported using an Al$_2$O$_3$ barrier [LYTV15].

On the fundamental point of view many questions about this system have still to be addressed. The first series of questions concern the SP: what should be the ideal stack (number of bilayers, Co and Ni thicknesses, end of the stack) in order to get a current passing through with the largest SP? How can large SP be obtained? Can we explain large SP looking at the electronic band structure? The second question is about the magnetic properties of this system. In a previous study we s that the Co magnetic moment (measured at room temperature) is enhanced at the Ni interface, and questions about possible measurement artefacts were addressed [GOTT12]. A complete X-Ray Magnetic Circular Dichroism (XMCD) study is proposed here to quantify the artefact due to the spin magnetic dipole operator at the interface. These new experiments also allowed us to measure the anisotropy of the orbital moment and compare it to the Bruno's PMA explanation [BRUN89]. The third question concerns the discrepancy between experimental and calculated Co atomic moment at the interface [GOTT12]. One explanation could be a special Co atomic arrangement at the interface with Ni not explored when performing *ab initio* calculations. To address these different points, we analyzed in details the electronic and structural properties of Co/Ni(111) superlattices grown by molecular beam epitaxy (MBE). In the first part, magnetic optical Kerr effect (MOKE) was used to determine the number of bilayers necessary to keep the PMA. In the second part Spin-Resolved PhotoEmission Spectroscopy (SR-PES) allowed us to show that close to full SP is achieved for 2 monolayer (ML) thick Co terminated stacks. In the third part X-Ray Magnetic Circular Dichroism (XMCD) led some new insight on the PMA origin in this system. In a fourth part, anomalous X-ray surface diffraction experiments were performed to address possible unusual atomic arrangement in the system. All these results are discussed in the fifth part.

## I – MOKE investigation

To definitely demonstrate that the perpendicular magnetic anisotropy is coming from the Co/Ni interface, stacks were grown with only one full Ni/Co first interface, with a varying Ni capping coverage as shown in fig.1a. The whole sample was then capped with MgO. The architecture of the sample is thus :

Au(111)/Ni$_{3ML}$/Co$_{xML}$/Ni$_{yML}$(wedge in y)/MgO

The hysteresis loops were measured along the Ni wedge by using Kerr microscopy applying the field perpendicular to the layers (fig.1b). As expected for a 3ML thick Co layer without Ni coverage, the out-of-plane axis was found to be a hard axis. Indeed, a unique Co/Ni interface anisotropy is not sufficient to overcome the demagnetization field. This is still the case up to 0.5 ML Ni cap, but the anisotropy (the loop slope) changed around 0.5 ML Ni cap and the out-of-plane axis became an easy axis above 0.7 ML Ni cap. Similar experiments were done for Co thickness varying from 1 to 4 ML. We deduced from these experiments the Ni cap thickness $y$ for which the magnetization easy axis turns from in-plane to out-of-plane for a Co layer thickness $x$ as plotted in fig.1c. The (x,y) couple of values may be determined by writing the total magnetic energy defining the effective anisotropy as ([JOHN96]):

$$K_{eff}D = \sum_{atoms\ i}(K_S + K_V^i t_i) + K_{shape}D \quad (1)$$

where $K_S$ are the interface magnetic anisotropy terms, $K_V^i$ the volumic magnetocrystalline anisotropy of the i$^{th}$ layer of thickness $t_i$, $D$ the total thickness and $K_{shape}$ the shape anisotropy terms. The transition from in-plane to out-of-plane easy axis corresponds to $K_{eff} = 0$ leading to a unique $(x,y)$ solution. The $(x,y)$ determination is however difficult first because the $K_{shape}$ term does not vary linearly with $x$ and $y$ and second because several interfaces are involved, i.e. Co/Ni, Co/MgO, Ni/MgO and Ni/Au (since the first Ni layer is grown on Au). However a rough estimation can be achieved by considering that $K_s^{Co/MgO} \approx K_s^{Ni/MgO}$ (noted $K_s^{MgO}$ in the following). Moreover, the Ni bulk anisotropy term $K_V^{Ni}$ is very small as shown by several groups [JOHN92, ZHAN93, GOTT12] and can be neglected. Considering our stack one can write:

$$K_{eff}D = K_s^{Co/Ni}(1+y) + K_s' + K_V^{Co}d.(x-1) + K_{shape}D \quad (2)$$

With $K_{shape}D = -\frac{\mu_0}{2}.\frac{(t_{Co}M_{Co}+t_{Ni}M_{Ni})^2}{t_{Co}+t_{Ni}}$,

$t_{Co} = d.x \quad t_{Ni} = d.(3+y)$ and $K_s' = K_s^{MgO} + K_s^{Ni/Au}$

Here $d \cong 0.2\ nm$ is the distance between atomic planes very similar in Ni and Co [GOTT12]. A second order equation on $y$ is obtained parametrized by $x$. Note that the bulk anisotropy term $K_V^{Co}$ is multiplied by $(x-1)$ since there is no bulk contribution for 1 Co atomic plane. In a previous study on a series of Co/Ni(111) superlattices, we measured $K_V^{Co} = +0.7 \pm 0.1\ MJ/m^3$ and $K_s^{Co/Ni} = +0.43 \pm 0.02\ mJ/m^2$ [GOTT12]. One should note here that both positive values help for PMA. This is the case for Co/Ni interface anisotropy and also "bulk" Co magnetocrystalline anisotropy since (111) is an easy axis. A very good fit is obtained (fig.1c) using $K_s' = -0.1 \pm 0.05\ mJ/m^2$. From [CHILD92] $K_s^{Ni/Au} = -0.15\ mJ/m^2$ was obtained in MBE-grown samples. This means that the MgO/Ni and MgO/Co interface anisotropies are small and not relevant here.

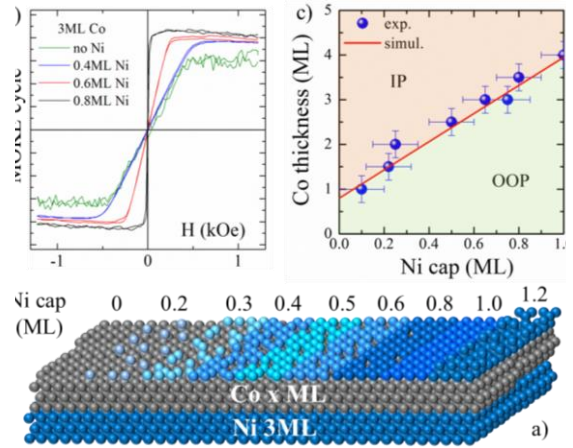

*Fig. 1. Co magnetic moment orientation depending on Ni capping measured by using Kerr magnetometry. a) Sample stack, b) Hysteresis loops for 3ML Co with different Ni capping, c) Ni cap thickness leading to in-plane (IP) to (OOP) out-of-plane magnetization transition for different Co thicknesses. The red curve is a fit using eq.(2).*

This analysis allowed us to estimate the number of bilayers necessary to get PMA under vacuum for photoemission experiments on Co terminated layer. For that $K_s^{Co/MgO}$ and $K_s^{Ni/MgO}$ should be changed in eq.(1) and (2) by $K_s^{Co/UHV}$ and $K_s^{Ni/UHV}$ given in ref.[JOHN96]. This analysis thus shows that 2 bilayers (Co$_x$Ni$_{3ML}$) without Ni cap on top are enough to get PMA up to x=2ML Co. However, for x=3ML a 3 repeats is necessary to get PMA. We actually verified this prediction when eprforming our spin polarized photoemission experiments.

## II – Co/Ni spin polarization

In order to investigate the SP of [Co/Ni] superlattices, SR-PES experiments were performed on the CASSIOPEE beamline at the SOLEIL synchrotron (see [ANDR14] for a description of the whole set-up). The Co/Ni stacks were epitaxially grown in a MBE chamber connected to the beamline to keep the surface pollution to an insignificant level [ANDR16]. We used the same growth process as developed in ref [GIRO09, GOTT12]. The films were deposited on $(11\bar{2}0)$ single-crystalline sapphire substrates. To deposit Co/Ni superlattices with (111) growth direction, thick seed layers of bcc V (110) / fcc Au (111) were first deposited. A series of [Ni(3ML)/Co(x)]$_3$ superlattices were measured with x ranging from 1 to 3 monolayers (ML) with 0.5ML step. Thick (0001) hcp Co and (111) fcc Ni films were also grown as bulk references. All Co/Ni superlattices have been magnetized with a 600 Oe field applied perpendicularly to the films before PES experiments (and in-plane for thick Ni and Co reference films). Preliminary experiments have shown that such a magnetic field is strong enough to saturate PMA Co/Ni magnetization [GIRO09]. The Spin-resolved photoemission experiments have been conducted at room temperature thanks to a spectrometer detector facing the sample surface with an angular integration to +/- 8°. The spectrometer is equipped with a Mott detector allowing to measure the spin polarization along in-plane and out-of-plane directions and thus allowing to check the magnetic anisotropy features of the Co/Ni stacks. Most of the experiments were performed using a 37 eV photon energy where the photoemission cross section is the largest for Co and Ni. Such conditions lead to probe around 40% of the first Brillouin Zone (BZ) in $k_x$-$k_y$ plane. To investigate the whole BZ, we thus turned the sample normal 8° off the detector axis. The PES experiments were performed on a (Co xML / Ni 3ML)*N superlattices series. The Ni thickness was fixed to y=3 ML (which is not a critical parameter since the PMA does not depend on it in our samples [GOTT12]). The Co thickness was varied from x=1 to 3 ML by 0.5 ML steps. The number of repetition N was fixed to 3 first to be sure to get the PMA according to the previous MOKE analysis and second because the PES probing depth is around 1 nm below the surface in our experimental conditions. In fig.2 the majority and minority spin PES and resulting SP for this series are plotted, including thick Ni and Co films (for which the magnetization was observed in-plane). The SP at the Fermi energy is much larger than in bulk Ni and Co and reaches 90% around 2ML Co. The SP increase is attributed to an increase of the minority spin PES near $E_F$ whereas the majority spin PES is weakly affected as indicated by the arrows in fig.2.

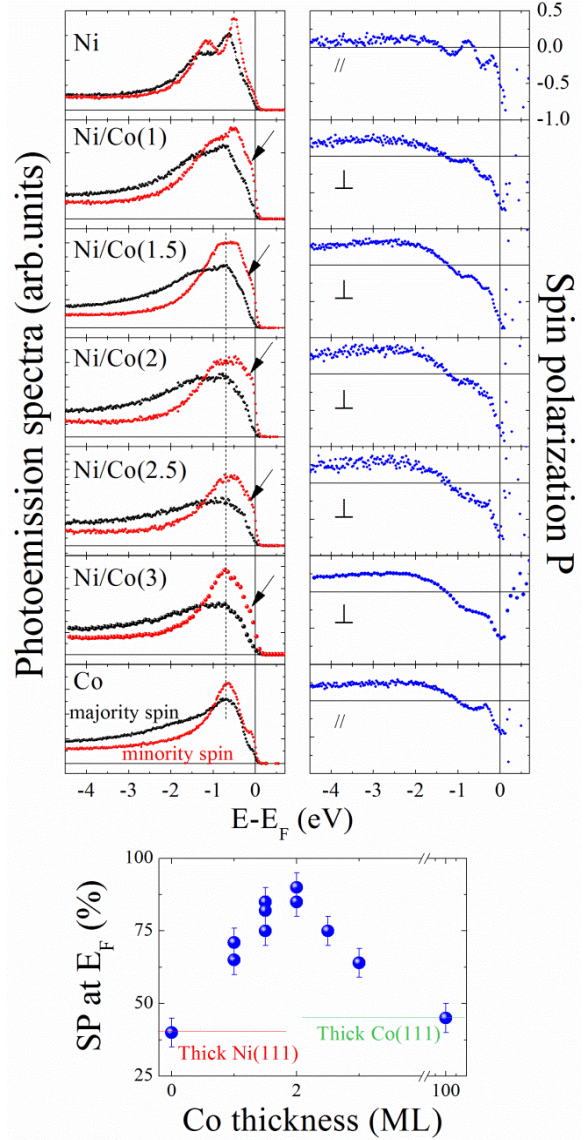

*Fig. 2. – top left- majority and minority spin PES spectra measured on bulk Ni and Co films and a series of Co/Ni superlattices, – top right- spin polarization spectra and – bottom- corresponding Spin Polarization (SP) at $E_F$. The arrows show the increase of the minority spin PES near $E_F$ responsible for the SP increase.*

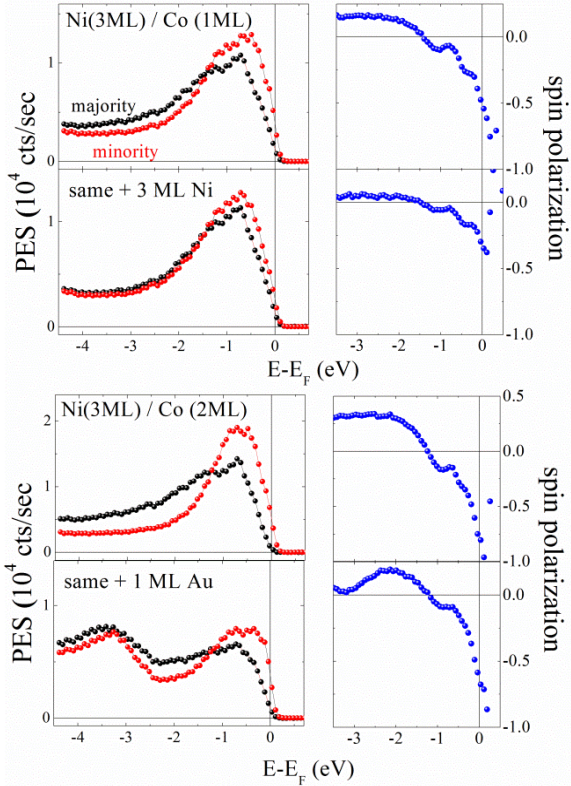

*Fig. 3. Effect of the covering on the spin-polarization of the stack, -top- for 3ML Ni capping and –bottom- for 1ML Au capping. This shows that Ni covering decreases the SP whereas an almost full polarized injection may be achieved in Au.*

We also looked at the impact of the capping on the Co layer. In fig.3 are shown the effect of Ni and Au covering on the SP at the surface. As expected the high SP obtained by a Co termination is reduced when covering with Ni. This leads to the conclusion that the use of Co/Ni SL as an electrode should be terminated with Co. The Au capping was also examined and the initial SP is slightly decreased. In practice the SP of the underneath Co layer is almost not affected because there is some unpolarised DOS coming from Au that is included in the calculation of the SP. This means that a very high spin polarised current can be injected in an Au spacer using a Co/Ni SL terminated with Co.

## III – PMA studied by XMCD

Using the XMCD technique we wanted to address some questions raised by a previous XMCD work [GOTT12] as: (i) what is the influence of the spin magnetic dipole operator on the spin moment determination using the sum rules and (ii) is the measurement of the orbital moment anisotropy $\Delta m_L$ comparable to the theoretical explanation proposed by Bruno [BRUN89]. In order to answer to these questions the absorption spectra were measured by at different angles between the sample and the photon beam, maintaining the saturation magnetization along the beam [STOH95]. These conditions require high magnetic fields to be fulfilled. The experiments were performed on the DEIMOS beamline at the SOLEIL synchrotron [OHRE14], [JOLY14]. The set-up is equipped with a superconducting coil allowing variable temperature (1.5-350 K) and high magnetic field measurements (up to 7T, we applied 4T here). Absorption spectra were recorded on a series of Ni (2ML) / Co (x ML) /Ni (3ML) /Au/V/Al$_2$O$_3$ samples (x=1, 2, 3) for 5 angles (noted γ, see inset in fig.4) between the photon / magnetic field direction and the (111) surface normal (γ=0, 15, 30, 45 and 60°). The raw XMCD spectra obtained at the Co edge for the 1ML Co sample are shown in fig.4. The orbital and effective spin moments were thus deduced using the sum rules and plotted versus $sin^2\gamma$ [WILH00]. All the absorption spectra were corrected from the saturation effect [NAKA99], [SICO05]. Such experiments allowed us to get the orbital moment anisotropy $\Delta m_L^i$ (i=Co, Ni) and the spin magnetic dipole operator $T_z$ contribution in the spin sum rule as [LAAN98]:

$$m_L = m_L^\perp + (m_L^\parallel - m_L^\perp)sin^2\gamma = m_L^\perp - \Delta m_L sin^2\gamma \quad (3)$$
$$m_s^{eff} = m_s + 7T_z = m_s + 14Q_{xx} - 21Q_{xx}sin^2\gamma \quad (4)$$

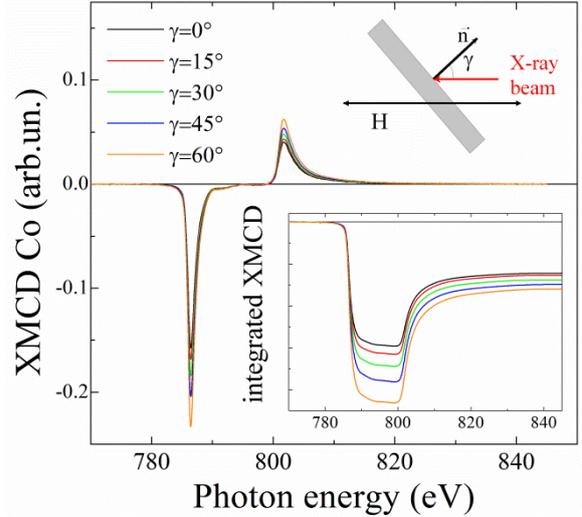

*Fig. 4. Examples of XMCD raw data observed at the Co L edge varying the angle γ between the X-ray beam and the surface normal. The magnetic field to get the XMCD is applied along the x-ray beam.*

Note that the 2$^{nd}$ equation is derived for a 3d metal with a uniaxial symmetry [DURR96] and shows that the anisotropy of $m_s^{eff}$ is only due to the anisotropy of $T_z$. The anisotropy of the orbital and spin effective moment are thus deduced from the slope of the curves $m_L(sin^2\gamma)$ and $m_s^{eff}(sin^2\gamma)$. These measurements were reproduced at 20, 50, 100, 200 and 300K. The results obtained on the 1ML Co sample are summarized in fig.5 and 6 for Co and Ni absorption edges respectively. From these results on the 1ML Co sample, we are first able to estimate the impact of the $T_z$ spin magnetic dipole operator on the Co and Ni spin moment determination using XMCD by looking at the slope of the $m_s^{eff}(sin^2\gamma)$ curves (see eq.3 and 4). This is observed negligible at all temperatures of measurement whereas a small but measurable $T_z$ effect is observed for Co. These two results are consistent with our previous *ab initio* calculation [GOTT12]. This experimental $T_z$ amplitude effect is also observed to increase when decreasing the temperature. It remains low even at 0K (less than 0.1 $\mu_B$/at) but enough large to take it into account for a correct $m_s$ determination. Second, the PMA is confirmed by the observation of a negative $m_L^\parallel - m_L^\perp$ slope with $sin^2\gamma$ for both Co and Ni atoms. We also observed that this orbital moment anisotropy is increasing when decreasing the temperature as expected. Similar experiments were performed for the 2ML and 3ML Co samples at room temperature. This XMCD analysis at the Co edge is shown in fig.7. The $T_z$ contribution is observed to almost vanish for 3ML Co which is not surprising since the $T_z$ contribution originates from the interfaces. We also observed that the orbital moment anisotropy decreases when increasing the Co thickness. Again this is consistent with the Co/Ni interface origin of the PMA.

These measurements also allow us to determine the gyromagnetic $g$-factor in both Ni and Co layers. This gives important information on the quality of the XMCD measurements and on the physics of PMA in this system [SHAW13] [AROR17]). Fig.8 displays the magnetization temperature dependence for the 1 and 3 ML thick Co samples. These magnetization variations are consistent with the reduced Curie temperature ($T_c$) of such stacks due first to size effects (small thicknesses) and second to different $T_c$ in bulk Co and Ni (1388 and 627K). Consequently the smallest is the Co thickness, the smallest the $T_c$ of the stack (for the same Ni quantity) and the higher the magnetization variation with temperature.

Finally, the $g$-factors can be extracted by using the equation:

$$\frac{\mu_l}{\mu_s} = \frac{g-2}{2} \quad (5)$$

Although the $g$ factor is an average value over the material as measured for instance by FerroMagnetic Resoannce (FMR) [SHAW13], one can extract atomic-like $g$ values from Ni and Co orbital and spin moments as plotted in fig.8. One should notice that these extracted $g$-factors are not dependent on temperature as expected. This reinforces the robustness of the measurement. Even if the accuracy on the $g$-factor is limited due to the addition of errors on Co and Ni moments, we find similar values for Ni and Co, meaning that these atomic-like $g$ factors are similar to those measured by FMR. These $g$-factors are quite far from the 2 value corresponding to free electrons. This is due to the strong Co/Ni hybridization at the interface also responsible for the PMA. We find similar $g$-factor values in Ni and Co around 2.25 for 1ML Co @300K. This is larger than the value measured for sputtered Co/Ni films, around 2.18 [SHAW13] [AROR17] yet this discrepancy can be readily explained. Indeed, we have shown that the PMA is slightly lower in sputtered than in MBE grown films [GOTT12] as was confirmed by other groups ([SHAW13, AROR17, SEKI17]). The present difference in $g$-factors is another signature of this PMA difference. Finally, it is interesting to compare the orbital moment anisotropies measured separately in Co and Ni. The anisotropy is 6 times larger in Co than in Ni (0.054$\mu_B$ and 0.009 $\mu_B$ respectively at 4K see fig.5 and 6) and one may consider that Ni plays a minor role in this process. Nevertheless, the ratio $m_L/m_s$ (proportional to the $g$-factor) for Co and Ni are similar in Co and Ni (fig.8).

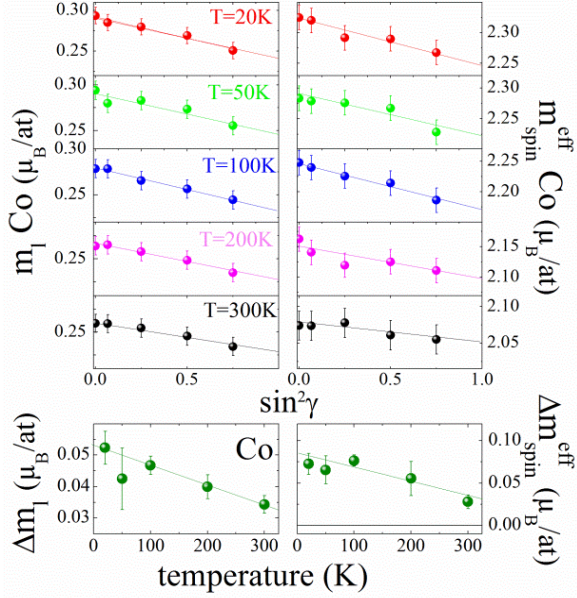

*Fig.5: -top- orbital and spin effective Co magnetic moments measured on the 1ML Co sample for different $\gamma$ angles and temperatures. -bottom- $\Delta m_L$ and $\Delta m_s^{eff}$ slopes plotted vs temperature.*

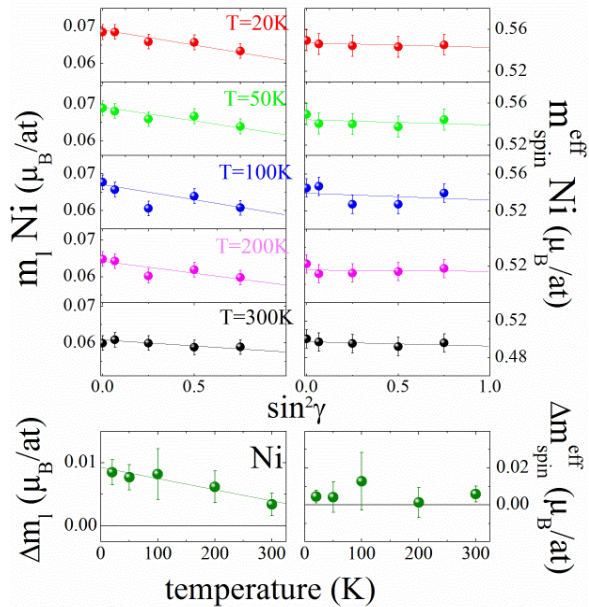

*Fig.6: similar experiments as in fig.5 but for Ni.*

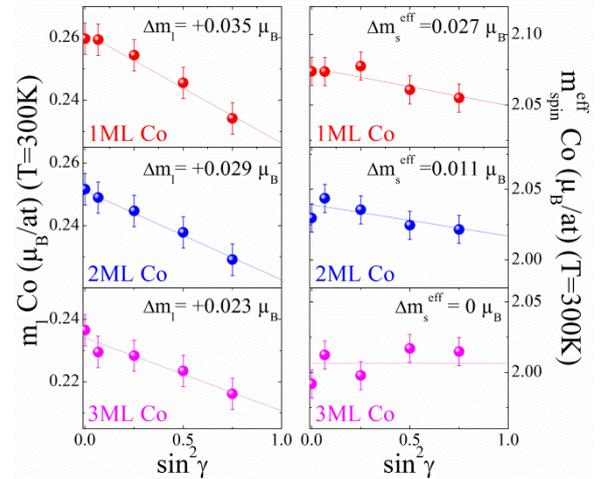

*Fig.7: Co orbital and spin effective moment anisotropies measured at room temperature for 3 samples with 1, 2 and 3ML Co. The Co PMA and the Tz contribution are observed to increase when decreasing the Co thickness. This is consistent with the interfacial origins of both anisotropies.*

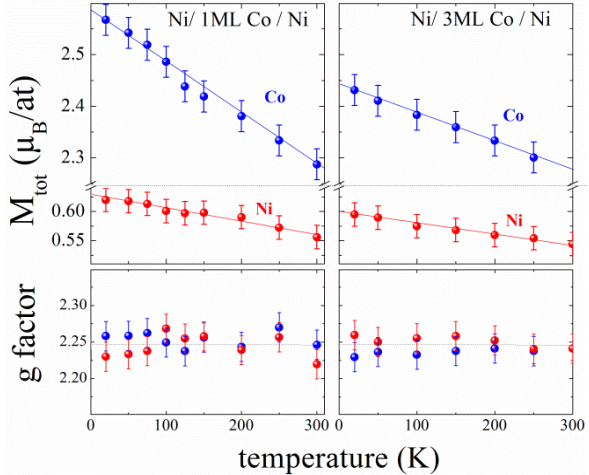

*Fig.8: -top- Total Co and Ni magnetic moment variation with temperature for the 1ML Co (left) and 3ML Co (right) samples –bottom- the g-factors extracted from these data.*

We can go further by looking at the macroscopic magnetic anisotropy link with the microscopic one in order to compare it to the Bruno's model. Our XMCD analysis allows us to extract the average $\Delta\mu_{orb}$ as:

$$\Delta\mu_l = \frac{t_{Co}\Delta\mu_l^{Co} + t_{Ni}\Delta\mu_l^{Ni}}{t_{Co} + t_{Ni}} \quad (6)$$

Moreover, we extracted the macroscopic anisotropy $K_u$ by measuring the effective anisotropy $K_{eff}$ on hysteresis loops performed by XMCD applying OOP and IP magnetic field (fig.9) and using:

$$K_u = K_{eff} + \frac{1}{2}\mu_o M_s^2 \quad (7)$$

According to Bruno's model, these macroscopic and microscopic anisotropies are linked as ([BRUN89], [SHAW13]):

$$K_u = A\frac{\xi N}{4V}\cdot\frac{\Delta\mu_l}{\mu_B} \quad (8)$$

Where $N/V$ is the atomic density (close to $8.9\ 10^{28}\ at/m^3$ here), $\xi$ the spin-orbit coupling parameter (we use the same value as in [SHAW13], $\xi \approx -1.6\times10^{-20}\ J/atom$), $\mu_B$ the Bohr magneton and $A$ a prefactor found in the literature to vary between 0.05 and 0.2 [WELL95, WILH00]. We report in fig.9 our data together with the results obtained on sputtered multilayer samples by Shaw's group [SHAW13]. We also confirm the Bruno's law but interestingly, we obtained a slope $A = 0.18$ around 2 times larger than those obtained on sputtered samples.

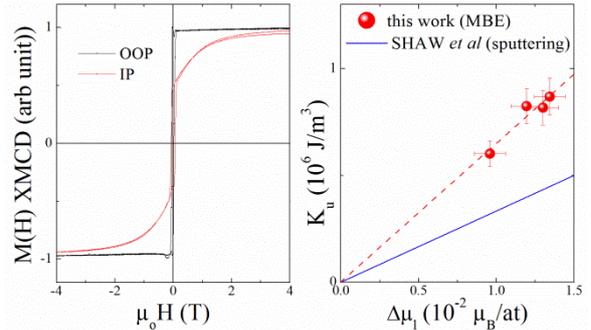

*Fig.9: -left- hysteresis loops measured by XMCD at the Co edge on a 1ML Co sample for Out-Of-Plane and In-Plane applied fields. The area between OOP and IP loops gives the effective anisotropy. –right- resulting macroscopic magnetic anisotropy versus the atomic orbital moment anisotropy. A linear law is found in agreement with Bruno's model (dashed red line). The blue line comes from results on sputtered samples reported in ref.[SHAW13].*

## IV – Crystalline arrangement studied by surface diffraction

The disagreements obtained between the *ab initio* magnetic moment calculations and the experimental results [GIMB11, GOTT12, SEKI17] motivated us to re-examine the crystallographic ordering in Co/Ni superlattices. Indeed, the calculations are carried out by considering a perfect FCC stack of hexagonal lattices, with distances between atoms similar or very close to those of the bulk phases. On the one hand, the distance between the Co and Ni planes at the interface may impact the electronic properties if, for example, it is much different as compared to bulk interplanar distance used in the calculations. Moreover, a particular arrangement of the Co and Ni atoms at the interface has to be considered. To get such information, we performed surface x-ray diffraction experiments on Co films epitaxially grown on a single-crystalline Ni(111) substrate. We also examined these deposits covered with a Ni monolayer.

In a surface diffraction experiment, the x-ray beam is sent at grazing incidence. The diffraction pattern consists in truncation rods perpendicular to the surface in the reciprocal space. A surface diffraction experiment consists in measuring these rods. Their profiles are extremely sensitive to various factors: atomic positions in the lattice, the type of atoms at each site, the stacking sequence, distances in the plane and out of the plane, and roughness [FEID89, ROBI91, RENA98]. These experiments generally require the use of synchrotron radiation given the strong reduction of the signal compared to conventional diffraction set-up. Moreover, the energy choice using Synchrotron radiation is mandatory in the present case since Ni and Co are neighbors in the periodic table so that their X-ray diffusion contrast is then very low using regular x-ray diffraction set-up. One elegant way of overcoming this difficulty is to set the x-ray beam energy at the 1s Co edge (hν = 7709eV). The Co diffusion factor is then strongly decreased by about 12 electrons whereas that of Ni remains close to 26 electrons at this energy, thus ensuring a high absorption contrast to X-rays (procedure called anomalous surface diffraction).

The experiments were carried out on the UHV diffraction station of the SixS beamline at the SOLEIL synchrotron source. The samples were prepared in the preparation chamber hosted by the diffractometer. The cleaning process of the substrate surface involved various cycles of Ar+ ion sputtering at 1 keV and subsequent annealing to 1100 K. The cleanliness and surface quality of the samples was checked by means of Auger electron spectroscopy and LEED, respectively. Co was deposited using an e-beam evaporator. The Co deposition rate was calibrated using a quartz microbalance located at the place of the sample. Co was deposited at room temperature (to avoid interdiffusion) at a rate of 1ML / min. For each sample the Ni(111) substrate was re-prepared following the above-mentioned procedure. Fig.10 a and b show the LEED pattern measured for the clean Ni(111) surface and after the Co deposition, respectively. Six different samples were prepared and analyzed: the bare Ni(111) substrate, x ML Co /Ni(111)substrate with x=1, 2 and 3, and 1ML Ni/ xML Co/Ni(111) with x=1 and 2.

The x-ray diffraction data presented in this paper are indexed according to the surface unit cell of Ni(111) described by the following parameters: $a = b = 0.2489\,nm$, $c = 0.609\,nm$ and $\alpha = \beta = 90°$, $\gamma = 120°$. Indeed, since Co can be grown in FCC (ABCABC stacking sequence) or in HCP (ABAB stacking), at least 4 planes stacked along (111) were considered to account for these two types of stacking (Fig.10c). A schematic view of the reciprocal space is represented in Fig.10d. For each of the above-mentioned samples we measured the integrated intensities along six crystal truncation rods on each sample: (11L), (1-2L), (-10L), (1-1L), (-11L), (0-1L) with L varying from 0.1 to 2.4 (larger values of L were not achievable due to geometric constrains). This gives 3 series of non-equivalent truncation rods along (fig.10):

(i) (11L), (1-2L) –Bragg peak at L=0,
(ii) (-10L),(1-1L) – Bragg peak at L=1
(iii) (-11L), (0-1L) – Bragg peak at L=2

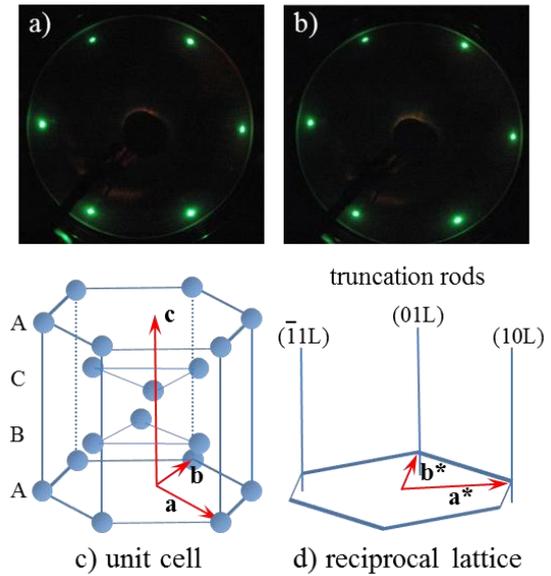

Fig.10: LEED patterns obtained on a) the Ni(111) surface and b) after the growth of 2ML thick Co film. The unit cell chosen for ROD simulation is indicated in c). Some truncation rods indexation is shown in d). 6 rods were measured for each sample.

After background subtraction and application of standard correction factors a fit of the data was carried out using the software package ROD [VILE00]. A detailed structural refinement was performed by fitting simultaneously three non-equivalent truncation rods. The experimental (blue circles) together with the best fit (red curve) structure factors are shown in Fig.11. During the refinement, the existence of two domains with different stacks had to be considered, whose percentages are also given on the right-hand side of Fig.11. It should be noted that in all the simulations the extracted roughness was found very close to zero (the simulations are shown for a roughness equal to zero). For the Ni substrate, during the refinement the only variable is the out-of-plane relaxation of the topmost plane. The simulations are excellent and almost no surface relaxation is observed (<1%) in agreement with the literature [WAN99].

For the 1ML Co film, the nucleation site of the Co layer must be taken into during the refinement. Considering the ABC stacking of the Ni substrate by terminating with a C plane, several options may be considered for Co nucleation: the nucleation site (A, B or C) is left free for both domains. The results of the refinement show that the site C is not occupied and that sites A (domain 1) and B (domain 2) can be occupied with some preference for site A. The fit also show that the Co film thickness is not exactly 1ML but 1.2 ML, which is within the error bar of the thickness control on this set-up. Therefore a second plane has grown on the first monolayer. Two types of stacking are obtained corresponding to 70% as an extension of the FCC structure of Ni and to 30% as a HCP stacking. When this type of deposit is covered with 1ML of Ni, the FCC stacking is then favored representing 90% of the surface.

The situation changes drastically for Co thickness higher than 1ML. For 2ML thick, the Co film adopts preferentially the HCP structure (domain 1, 70%), even if there remain zones with an FCC stacking (domain 2, 30%). For 3ML Co, the HCP stacking dominates and no more FCC stacking occurs. The Ni capping no longer affects the stacking of the Co which remains HCP, contrary to the previous case with a single plane of Co. Finally, the interplanar distances for all these samples obtained following the data refinement are found very close to the bulk Ni atomic distance. The out-of-plane relaxations never exceed +/- 3%, which is of the order of magnitude of the misfit between Co and Ni bulk lattices.

Therefore this analysis clearly demonstrates that there is no special atomic arrangement other than the FCC or HCP stacking, and no strong lattice parameter relaxation. The only parameter that changes from one sample to another is the stacking sequence. For 1ML Co film the FCC stacking is preferred, but the HCP stacking dominates for Co films thicker than 2ML. This means that an atomic plane of Co needs to be covered by Co to generate the HCP structure.

## V – *Ab initio* calculations

The electronic structure of the Ni(3MLs) / Co(2Mls) superlattices with different stacking of the Ni and Co monolayers have thus been calculated from first principles, using the code Wien2k [BLAH90] and the Perdew-Burke-Ernzerhof (PBE) [PERD96] functional for describing the exchange and correlation potential. We used atomic spheres with a radius of 0.1164 nm (2.2 atomic units), an in-plane lattice parameter of 0.2507 nm and a distance of 0.2047 nm between successive atomic layers (measured on our samples [GOTT12]. We considered 4 different stacking, which differ by the sequences of the successive Ni and Co monolayers occupying the conventional A, B, or C atomic sites of the fcc lattice. The 4 superlattices that we studied correspond to:

(i) a perfect fcc stacking, labelled "Ni(fcc) / Co(fcc)" with the sequence Ni(ABC) / Co(AB) / Ni(CAB) / Co(CA) / Ni(BCA) / Co(BC),
(ii) a first stacking combining fcc Ni and hcp Co, labelled "Ni(fcc)/Co(hcp)1", with the sequence Ni(ABC) / Co(BC) / Ni(BCA) / Co(CA) / Ni(CAB) / Co(AB),
(iii) a second stacking combining fcc Ni and hcp Co, labelled "Ni(fcc)/Co(hcp)2", with the sequence Ni(ABC) / Co(AC),
(iv) finally a perfect hcp stacking, labelled "Ni(hcp)/Co(hcp)", with the sequence Ni(ABA) / Co(BA) / Ni(BAB) / Co(AB).

The spin magnetic moments that we calculated for the different atoms in these SLs are given in Table I. To summarize, the stacking sequence has a small influence on the averaged nickel magnetic moment and on the Co spin magnetic moment, which only vary by 2.6% and 0.5%, respectively. Finally, we also varied the distance at the interface between Co and Ni up to 5% and the Co atomic moment was observed to vary insignificantly compared to our observations.

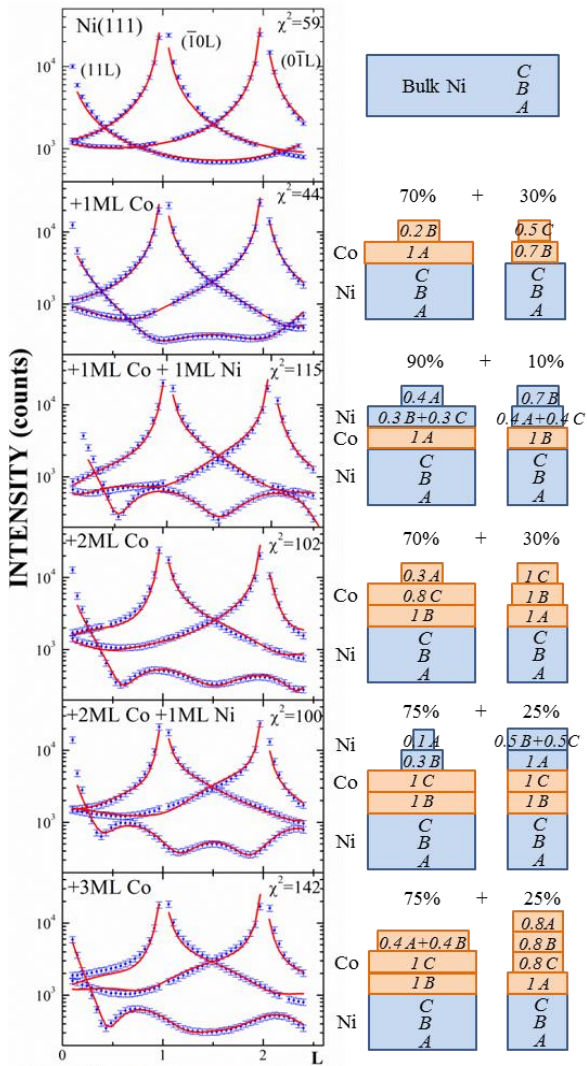

*Fig.11: experimental structure factors (blue open circles, together with error bars) and the best fit structure factors (solid red line) truncations rods and simulation for the six samples, from top to bottom: Ni(111) substrate, Ni + 1ML Co, Ni + 1ML Co+ 1ML Ni, Ni + 2ML Co, Ni + 2ML Co + 1ML Ni, Ni + 3ML Co. The schematic representation of the models of stacks obtained after ROD simulations is shown on the right.*

| Superlattices stacking | Spin magnetic moment (Bohr magneton) | | | |
|---|---|---|---|---|
| | Ni interface | Ni central | Ni average | Co |
| Ni(fcc)/Co(fcc) | 0.6784 | 0.6387 | 0.6651 | 1.7426 |
| Ni(fcc)/Co(hcp)1 | 0.6994 | 0.5997 | 0.6662 | 1.7423 |
| Ni(fcc)/Co(hcp)2 | 0.6983 | 0.6174 | 0.6713 | 1.7493 |
| Ni(fcc)/Co(hcp) | 0.7091 | 0.6289 | 0.6824 | 1.7510 |

*Table I: Calculated spin magnetic moment of Ni and Co atoms at the interfaces and of Ni atoms at the center of the Ni layers in Ni(3MLs)/Co(2MLs) superlattices with different stacking sequences. The averaged value of the nickel spin atomic moment is also given.*

## VI – Discussion

The first point we want to address concerns the link between macroscopic anisotropy via $K_u$ and microscopic anisotropy via $\Delta m_l$. The agreement with the Bruno's model is quite comparable to that observed in other epitaxial systems. However, this law is not similar for epitaxial Co/Ni superlattices (MBE) or for multilayers (sputtering). This observation is not surprising since at least 4 different groups [GOTT12, SHAW13, AROR17, SEKI17] came to the same conclusion: the macroscopic magnetic anisotropy $K_u$ is higher in epitaxial films. However, the anisotropy on the atomic orbital moment ($\Delta m_l$) observed here is similar to the values reported by Shaw and co-workers [SHAW13]. This strongly suggests that the origin of the $K_u$ difference obtained between epitaxial and sputtered samples does not occur at the atomic level but on a larger scale (roughness ? small impurities concentration?).

Concerning the atomic magnetic moments, the spin moment of Ni is therefore very close to the bulk value, whereas a noticeable increase of the orbital moment is observed here compared to the bulk. The situation is more complex for Co. The result of the surface diffraction study indicates that two different sites (and only two) need to be considered in the stack of Co layers. At the interface with Ni, the 2 contributions of spin and orbit are strongly increased with respect to bulk (HCP or FCC) Co values. On the other hand, for the Co atoms surrounded by Co (the central plane for a layer of 3 planes for example), the measured moments are then comparable to the bulk values. The strong increase of the moments at the interface with Ni had previously been discussed as possibly coming from the application of the sum rules in particular for the spin moment determination. Indeed, the contribution of $T_z$ (which can strongly increases at a symmetry breaking interface) could affect the estimation of the spin moment. In this study, it was possible to determine experimentally this contribution. We find it to be low even if it is not negligible. The "true" values of the spin moments can thus be determined. However, they are still increased by about 30% compared to a whole Co environment, whereas a 6% increase is given by the calculations considering a perfect FCC structure of superlattices [GOTT12]. We have therefore carried out calculations by considering an HCP stack of Co. The results remain unchanged. Finally, we have varied in the calculations the distances between the plane of Co at the interface with Ni and in the Co layers of several %, in any case beyond what we determined using anomalous surface diffraction. Again, we did not get any increase in Co moments as observed experimentally.

Consequently, there is a flagrant disagreement between the Co moments determined experimentally and the calculated ones by considering a crystallographic structure in agreement with the observation given by diffraction. This disagreement is surprising since a very simple model gives a qualitative explanation for the observations. Indeed, one can consider on the one hand that at the interface Co and Ni exchange electrons in such a way that the 3d band of Co is filled with 7.5 electrons [GIMB11, GOTT12]. On the other hand, if we consider that the 3d band of the majority electrons is completely filled at the interface, there remain 2.5 electrons in the minority band. This scenario then leads to a full spin polarization and to an atomic moment of 2.5 $\mu_B$ for Co at the interface with Ni, in agreement with the experiment.

## VII – Conclusion

In this paper, we brought a new experimental proof that the PMA in Co/Ni superlattices is closely linked to the Co/Ni interface. Interestingly, the effective magnetic anisotropy can be tuned by controlling the Ni coverage on a Co film. Such a possibility may be an opportunity for switching the magnetization by an external electric field (in a tunnel barrier for instance), using an adequate Ni coverage to get the system very close to the in-plane to out-of-plane transition. The angle-dependent XMCD analysis using strong field and low temperature allowed us to confirm the very high magnetic moment of Co in contact with Ni (2.55$\mu_B$ to be compared to the bulk value 1.7 $\mu_B$). Such a strong enhancement has still to be reproduced using *ab initio* calculations but our surface diffraction results drastically limit the number of configurations that have to be considered. The anisotropy of the orbital magnetic moment is shown to be linked to the magneto-crystalline anisotropy following the Bruno's model. As a consequence, the much stronger orbital moment anisotropy in Co compared to Ni demonstrates that PMA is essentially due to Co. This analysis also sheds some light on the PMA difference observed in sputtered and MBE-grown samples. The anisotropy at the atomic level ($\Delta\mu_l$) are similar using both growth techniques, whereas the macroscopic anisotropy ($K_u$) is lower on sputtered films. This strongly suggests that the origin of this magnetic anisotropy difference is not at the atomic level but on a large scale. Finally, the most interesting result in this study regarding application in STT-based devices is the very high spin polarization obtained on Co/Ni stacks terminated by Co, reaching at least 90% for 2 ML Co. These PMA and high spin polarization behaviors are essential to build efficient devices based on spin transfer torque.

*Acknowledgement*
This work was supported partly by the french PIA project "Lorraine Université d'Excellence", reference ANR-15-IDEX-04-LUE. The theoretical calculations were granted access to the HPC resources of CALMIP supercomputing center under the allocation P1252.

**References:**

[SANK07] *Measurement of the Spin-Transfer-Torque Vector in Magnetic Tunnel Junctions* J.C. Sankey, C, Y-T Cui, J.Z. Sun, J.C. Slonczewski, R.A. Buhrman, D.C Ralph, *Nat. Phys.* **4**, 67 (2007)

[MEEN14] *Overview of Emerging Nonvolatile Memory Technologies* J.S. Meena, M.S. Simon, C. Umesh, T-Y Tseng, *Nanoscale Res. Lett.* **9**, 526 (2014)

[SILV10] *Developments in nano-oscillators based upon spin-transfer point-contact devices* T. Silva, and W. Rippard, J. Magn. Magn. Mater. **320**, 1260–1271 (2010)

[DUME12] *Molecular Beam Epitaxy: From Quantum Wells to Quantum Dots. From Research to Mass Production* Chap 20 : *Epitaxial Magnetic Layers Grown by MBE : Model Systems to Study the Physics in Nanomagnetism and Spintronic* K. Dumesnil & S. Andrieu, Ed. M. Henini, ELSEVIER (2012)

[RAJA10] *Spin polarization of currents in Co/Pt multilayer and Co–Pt alloy thin films* A. Rajanikanth, S. Kasai, N. Ohshima, K. Ohno, *Appl. Phys. Lett*. **97**, 022505 (2010)

[SAKU06] *Giant tunneling magnetoresistance in Co2Mn Si / Al − O / Co2MnSi magnetic tunnel junctions* Y. Sakuraba, M. Hattori, M. Oogane, Y. Ando, H. Kato, A. Sakkuma, T. Miyazaki, H. Kubota, *Appl. Phys. Lett*. **88**, 192508 (2006)

[MARU06] *High tunnel magnetoresistance in fully epitaxial magnetic tunnel junctions with a full-Heusler alloy Co2Cr0.6Fe0.4Al thin film* T. Marukame, T. Ishikawa, K. Matsuda, T. Uemura, and M. Yamamoto, *Appl. Phys. Lett*. **88**, 262503 (2006)

[TEZU06] *175% tunnel magnetoresistance at room temperature & high thermal stability using $Co_2FeAl_{0.5}Si_{0.5}$ full-Heusler alloy electrodes* N. Tezuka, N. Ikeda, S. Sugimoto, K. Inomata, *Appl. Phys. Lett*. **89**, 252508 (2006)

[ANDR16] *Direct evidence for minority spin gap in the Co2MnSi Heusler compound* S. Andrieu, A. Neggache, T. Hauet, T. Devolder, A. Hallal, M. Chshiev, A.M. Bataille, P. Le Fevre, F. Bertran, *Phys. Rev. B* **93**, 094417 (2016)

[PRAD17] *First-principles calculation of the effects of partial alloy disorder on the static and dynamic magnetic properties of $Co_2MnSi$*, B. Pradines, R. Arras, I. Abdallah, N. Biziere and L. Calmels, *Phys Rev B* **95**, 094425 (2017)

[JOHN96] *Magnetic anisotropy in metallic multilayers* M.T. Johnson, P.J.H Bloemen, F.J.A den Broeder, J.J. de Vries, *Rep. Prog. Phys.* **59**, 1409 (1996)

[ANDE06] *Perpendicular Magnetocrystalline Anisotropy in Tetragonally Distorted Fe-Co Alloys* G. Andersson, T. Burkert, P. Warnicke, M. Björck, B. Sanyal, C. Chacon, C. Zlotea, L. Nordström, P. Nordblad, O. Eriksson, *Phys. Rev. Lett.* **96**, 037205 (2006)

[OUAZ12] *Atomic-scale engineering of magnetic anisotropy of nanostructures through interfaces and interlines* S. Ouazi, S. Vlaic, S. Rusponi, G. Moulas, P. Buluschek, K. Halleux, S. Bornemann, S. Mankovsky, J. Minar, J.B. Staunton, H. Ebert & H. Brune, *Nat. Com.* **3**, 1313 (2012)

[YOU12] *Co/Ni multilayers with perpendicular anisotropy for spintronic device applications* L. You, R.C. Sousa, S. Bandiera, B. Rodmacq, B. Dieny, *Appl. Phys. Lett.* **100**, 172411 (2012)

[GOTT12] *Co/Ni(111) superlattices studied by microscopy, X-ray absorption and ab-initio calculations* M. Gottwald, S. Andrieu, F. Gimbert, E. Shipton, L. Calmels, C. Magen, E. Snoeck, M. Liberati, T. Hauet, E. Arenholz, S. Mangin, E.E. Fullerton, *Phys. Rev. B* **86**, 014425 (2012)

[JOHN92] *Orientational Dependence of the Interface Magnetic Anisotropy in Epitaxial Ni/Co/Ni* M.T. Johnson, J. J. de Vries, N. W. E. McGee, J. aan de Stegge, F.J.A. den Broeder, *Phys. Rev. Lett.* **69**, 3575 (1992)

[ZHAN93] *Anisotropy and Magneto-Optical Properties of Sputtered Co/Ni Multilayer Thin Films* Y. B. Zhang, J. A. Woollam, Z. S. Shan, J. X. Shen, and D. J. Sellmyer, *IEEE Trans. Magn.* **30**, 4440 (1994)

[GIRO09] *Strong perpendicular magnetic anisotropy in Ni/Co(111) single crystal superlattices* S. Girod, M. Gottwald, S. Andrieu, S. Mangin, J. McCord, Eric E. Fullerton,J.-M. L. Beaujour, B. J. Krishnatreya, A. D. Kent, *Appl. Phys. Lett.* **94**, 262504 (2009)

[SHAW13] *Measurement of orbital asymmetry and strain in Co90Fe10/Ni multilayers and alloys: Origins of perpendicular anisotropy* J. M. Shaw, H. T. Nembach, T. J. Silva, *Phys. Rev. B* **87**, 054416 (2013)

[DAAL92] *Prediction and confirmation of perpendicular magnetic anisotropy in Co/Ni multilayers* G.H.O. Daalderop, P.J. Kelly, F.J.A. den Broeder, *Phys. Rev. Lett.* **68**, 682 (1992)

[KYUN96] *Perpendicular magnetic anisotropy of metallic multilayers composed of magnetic layers only/ Ni/Co and Ni/Fe multilayers* K. Kyuno, J. G. Ha, R. Yamamoto, and S. Asano, *Jpn. J. Appl. Phys.* **35**, 2774 (1996).

[GIMB12] *First-principles investigation of the magnetic anisotropy and magnetic properties of Co/Ni(111) superlattices* F. Gimbert and L. Calmels, *Phys. Rev. B* **86**, 184407 (2012)

[BEAU07] *Ferromagnetic resonance study of sputtered Co|Ni multilayers* J.-M. Beaujour, W. Chen, K. Krycka, C.-C. Kao, J.Z. Sun, A.D. Kent, *Eur. Phys. J. B* **59**, 475 (2007)

[CHEN08] *Spin-torque driven ferromagnetic resonance of Co/Ni synthetic layers in spin valves* W. Chen, J.-M.L. Beaujour, G. deLoubens, A.D. Kent, *Appl. Phys. Lett.* **92**, 012507 (2008)

[SEKI17] *Magnetic Anisotropy and Damping for Monolayer-Controlled Co|Ni Epitaxial Multilayer* T. Seki, J. Shimada, S. Iihama, M. Tsujikawa, T. Koganezawa, A. Shioda, T. Tashiro, W. Zhou, S. Mizukami, M. Shirai, K. Takanashi, *J. Phys. Soc. Jpn. 86, 074710 (2017)*

[MANG06] *Current-induced magnetization reversal in nanopillars with perpendicular anisotropy* S. Mangin, D. Ravelosona, J.A. Katine, M.J. Carey, B.D. Terris, E.E. Fullerton, *Nat. Mater.* **5**, 210 (2006)

[MANG09] *Reducing the critical current for spin-transfer switching of perpendicularly magnetized nanomagnets* S. Mangin, Y. Henry, D. Raveloson, J.A. Katine, E.E. Fullerton, *Appl. Phys. Lett.* **94**, 012502 (2009)

[BERN11] *Nonuniform switching of the perpendicular magnetization in a spin-torque-driven magnetic nanopillar* D.P. Bernstein, B. Brauer, R. Kukreja, J. Stohr, T. Hauet, J. Cucchiara, S Mangin, J.A. Katine, T. Tyliszczak, K.W. Chou, Y. Acremann, *Phys. Rev. B* (R) **83**, 180410 (2011)

[TANI09] *Domain Wall Motion Induced by Electric Current in a Perpendicularly Magnetized Co/Ni Nano-Wire* H. Tanigawa, T. Koyama, G. Yamada, D. Chiba, S. Kasai, S. Fukami, T. Suzuki, N. Ohshima, N. Ishiwata, *Appl. Phys. Exp.* **2**, 053002 (2009)

[UEDA12] *Temperature dependence of carrier spin polarization determined from current-induced domain wall motion in a Co/Ni nanowire* K. Ueda, T. Koyoma, R. Hiramatsu, D. Chiba, S. Fukami, H. Tanigawa, T. Suzuki, N. Ohshima, N. Ishiwata, Y. Nakatani, K. Kobayashi, T. Ono, *Appl. Phys. Lett.* **100**, 202407 (2012)

[GIMB11] *Localized electron states & spin polarization in Co/Ni(111) overlayers* F. Gimbert, L. Calmels, S. Andrieu, *Phys. Rev. B* **84**, 094432 (2011)

[LEGA15] *Thermally activated domain wall motion in [Co/Ni](111) superlattices with perpendicular magnetic anisotropy* S. Legal, N. Vernier, F. Montaigne, M. Gottwald, D. Lacour, M. Hehn, J. Mc Cord, D. Ravelosona, S. Mangin, S. Andrieu, T. Hauet *Appl. Phys. Lett.* **106**, 062406 (2015)

[LEGA17] *Effect of spin transfer torque on domain wall motion regimes in [Co/Ni] superlattice wires* S. Legal, N. Vernier, F. Montaigne, A. Thiaville, J. Sampaio, D. Ravelosona, S. Andrieu, T. Hauet *Phys. Rev. B* **95**, 184419 (2017)

[LYTV15] *Magnetic tunnel junctions using Co/Ni multilayer electrodes with perpendicular magnetic anisotropy* Ia. Lytvynenko, C. Deranlot, S. Andrieu, T. Hauet, *J. Appl. Phys.* **117**, 053906 (2015)

[BRUN89] *Tight-binding approach to the orbital magnetic moment and magnetocrystalline anisotropy of transition-metal monolayers* P. Bruno, *Phys. Rev. B* **39**, 865 (1989)

[KAMA99] *Structure & magnetic properties of Au/Ni/Ag & Ag/Ni/Au superlattices* Y. Kamada, H. Kasai, T. Kingetsu, M. Yamamoto, *J. Magn. Soc. Japan* **23**, 581 (1999)

[YANG11] *First-principles investigation of the very large perpendicular magnetic anisotropy at Fe|MgO & Co|MgO interfaces* H. X. Yang, M. Chshiev, B. Dieny, J. H. Lee, A. Manchon, K. H. Shin, *Phys. Rev. B* **84**, 054401 (2011)

[LAM13] *Composition Dependence of Perpendicular Magnetic Anisotropy in Ta/CoxFe80-xB20/MgO/Ta (x=0, 10, 60) Multilayers* D. D. Lam, F. Bonell, S. Miwa, Y. Shiota, K. Yakushiji, H. Kubota, T. Nozaki, A. Fukushima, S. Yuasa, Y. Suzuki, *J. of Magnetics* **18**, 5 (2013)

[ANDR14] *Spectroscopic & transport studies of CoxFe1-x/MgO based MTJs* S. Andrieu, L. Calmels, T. Hauet, F. Bonell, P. Le Fevre, F. Bertran, *Phys. Rev. B* **90**, 214406 (2014)

[STOH95] *Determination of Spin- and Orbital-Moment Anisotropies in Transition Metals by Angle-Dependent X-Ray Magnetic Circular Dichroism*
J. Stöhr and H. König, *Phys. Rev. Lett.* **75**, 3748 (1995)

[OHRE14] *DEIMOS: A beamline dedicated to dichroism measurements in the 350–2500 eV energy range*
P. Ohresser, E. Otero, F. Choueikani, K. Chen, S. Stanescu, F. Deschamps, T. Moreno, F. Polack, B. Lagarde, J-P. Daguerre, F. Marteau, F. Scheurer, L. Joly, J-P. Kappler, B. Muller, O. Bunau, P. Sainctavit,
*Review of Scientific Instruments*, **85**, 013106 (2014)

[JOLY14] *Fast continuous energy scan with dynamic coupling of the monochromator and undulator at the DEIMOS beamline*
L. Joly, E. Otero, F. Choueikani, F. Marteau, L. Chapuis, P. Ohresser, *Journal of Synchrotron Radiation* **21**, 502 (2014)

[LAAN98] *Microscopic origin of magnetocrystalline anisotropy in transition metal thin films*
G. van der Laan, *J. Phys. Cond. Mat.* **10**, 3239 (1998)

[WELL95] *Microscopic Origin of Magnetic Anisotropy in Au/Co/Au Probed with X-Ray Magnetic Circular Dichroism*
D. Weller, J. Stohr, R. Nakajima, A. Carl, M. G. Samant, C. Chappert, R. Megy, P. Beauvillain, P. Veillet, G. A. Held, *Phys. Rev. Lett.* **75**, 3752 (1995)

[WILH00] *Magnetic anisotropy energy and the anisotropy of the orbital moment of Ni in Ni/Pt multilayers*
F. Wilhelm, P. Poulopoulos, P. Srivastava, H. Wende, M. Farle, K. Baberschke, M. Angelakeris, N. K. Flevaris, W. Grange, J.-P. Kappler, G. Ghiringhelli, and N. B. Brookes, *Phys. Rev. B* **61**, 8647 (2000)

[NAKA99] *Electron-yield saturation effects in L-edge x-ray magnetic circular dichroism spectra of Fe, Co, and Ni*
R. Nakajima, J. Stohr, and Y. U. Idzerda, *Phys. Rev. B* **59**, 6421 (1999)

[SICO05] *Electronic properties of Fe, Co, and Mn ultrathin films at the interface with MgO(001)* M. Sicot, S. Andrieu, F. Bertran, F. Fortuna, *Phys. Rev B* **72**, 144414 (2005)

[AROR17] *Origin or perpendicular magnetic anisotropy in Co/Ni multilayers* M. Arora, R. Hübner, D. Suess, B. Heinrich and E. Girt, *Phys. Rev. B* **96**, 024401 (2017)

[DURR96] *Magnetic circular x-ray dichroism in transverse geometry: Importance of noncollinear ground state moments* H.A. Dürr and G. van der Laan. *Phys. Rev. B* **54**, R760 (1996)

[FEID89] *Surface structure determination by X-ray diffraction* R. Feidenhans'l, *Surf.Sci. Reports* **10**, 105 (1989)

[ROBI91] I.K. Robinson, *Surface Crystallography*, Chap. 7 in Handbook on Synchrotron Radiation, Vol. 3, eds. G. Brown & D.E. Moncton (North-Holland, Amsterdam, 1991)

[RENA98] *Oxide surfaces and metal/oxide interfaces studied by grazing incidence X-ray scattering*
G. Renaud, *Surf. Sci. Reports* **32**, 1 (1998)

[VLIE00] *ROD, a program for surface X-ray crystallography* E. Vlieg, *J. Appl. Cryst.* **33**, 401 (2000)

[WAN99] *Surface relaxation and stress of fcc metals: Cu, Ag, Au, Ni, Pd, Pt, Al and Pb*
J. Wan, Y L Fan, D W Gong, S G Shen and X Q Fan, *Modelling Simul. Mater. Sci. Eng.* **7** 189 (1999)

[BLAH90] *An Augmented Plane Wave + Local Orbitals Program for Calculating Crystal Properties* P. Blaha, K. Schwarz, P. Sorentin, S.B. Trickey, *Comp. Phys. Com.* **59**, 399 (1990)

[PERD96] *Generalized Gradient Approximation Made Simple* J.P. Perdew, K. Burke, M. Ernzerhof, *Phys. Rev. Lett.* **77**, 3865 (1996)